\documentclass[twocolumn,superscriptaddress,aps,preprintnumbers,amsmath,amssymb,prl,nofootinbib
]{revtex4}

\usepackage{graphicx}
\usepackage{epstopdf}
\usepackage{dcolumn}
\usepackage{bm}
\usepackage{hyperref}
\usepackage{color}

\usepackage{amsmath}

\def\dd{\mathrm{d}}

\def\em{{\rm em}}

\renewcommand{\emph}[1]{\textit{#1}}

\begin{document}

\title{Axion Dark Matter Search with Interferometric Gravitational Wave Detectors}

\author{Koji Nagano}
\affiliation{Institute for Cosmic Ray Research, University of Tokyo, Kashiwa 277-8582, Japan}
\author{Tomohiro Fujita}
\affiliation{Department of Physics, Kyoto University, Kyoto, 606-8502, Japan}
\affiliation{D\'epartment de Physique Th\'eorique and Center for Astroparticle Physics, \\ Universit\'e de Gen\`eve, Quai E.Ansermet 24, CH-1211 Gen\`eve 4, Switzerland}
\author{Yuta Michimura}
\affiliation{Department of Physics, University of Tokyo, Bunkyo, Tokyo 113-0033, Japan}
\author{Ippei Obata}
\affiliation{Institute for Cosmic Ray Research, University of Tokyo, Kashiwa 277-8582, Japan}

\begin{abstract}
Axion dark matter differentiates the phase velocities of the circular-polarized photons.
In this Letter, a scheme to measure the phase difference by using a linear optical 
cavity is proposed. If the scheme is applied to the Fabry-P\'erot arm of Advanced LIGO-like (Cosmic-Explorer-like) gravitational wave
detector, the potential sensitivity to the axion-photon coupling constant, $g_{\text{a}\gamma}$, reaches $g_{\text{a}\gamma} \simeq 8\times10^{-13} 
\text{~GeV}^{-1}\, (4 \times 10^{-14}\text{~GeV}^{-1})$  
at the axion mass $m \simeq 3\times 10^{-13}$ eV ($2\times10^{-15}$ eV)
and remains at around this sensitivity for 3 orders of magnitude in mass.
Furthermore, its sensitivity has a sharp peak reaching 
$g_{\text{a}\gamma} \simeq 10^{-14} \text{~GeV}^{-1}\  (8\times10^{-17} \text{~GeV}^{-1})$ 
at $m = 1.563\times10^{-10}$ eV ($1.563\times10^{-11}$ eV).
This sensitivity can be achieved without loosing any sensitivity to gravitational waves.

\end{abstract}




\maketitle

%
%
%
\section{Introduction}

Axion is a pseudo-scalar field that is originally proposed in the late 1970s 
to solve the strong CP problem in QCD physics, known as ``QCD axion"~\cite{Peccei1977}.
In recent decades, it has been found that high energy physics such as string theory also predicts a number of axion-like particles
from the compactification of extra dimensions~\cite{Svrcek2006}.
Hereafter we collectively call them ``axion".
Axion typically has a small mass $m \ll \text{eV}$ 
and behaves like non-relativistic fluid in the present universe due to its oscillatory behavior.
For this reason, axion is a cosmologically well-motivated candidate of dark matter. Another important feature of axion is its coupling to gauge bosons.
In particular, a small but finite coupling between axion and photon is a general prediction of high energy physics and it provides  a good chance to detect axion by using the well-developed photonics technology. 

The conventional way to probe axion is to look for a phenomena that
axion and photon are converted each other under the background magnetic field, known as the axion-photon conversion \cite{Sikivie1983}.
Many experiments and astronomical observations have been performed to probe axion  via the axion-photon conversion \cite{Hagmann1998, CASTCollaboration2005, Vogel2013, Ehret2009, Betz2013, Tam2012, Sikivie2014, Kahn2016, Silva-Feaver2017, Brockway1996, Payez2015, Wouters2012, Marsh2017, Conlon2018, Aharonian2007, Conlon2013, Kohri:2017ljt, Moroi2018, Caputo2019}
, while no signal has been found (for recent reviews, see \cite{Dias2014}).
Recently, however, a new experimental approach to search for axion dark matter was proposed which does not need a strong magnetic field but uses optical cavity~\cite{Melissinos2009, DeRocco2018, Obata2018, Liu2018}.
This new method aims to measure the difference of phase velocity between two circular-polarized photons which is caused by the coupling to axion dark matter~%
\cite{Carroll1990, Andrianov2010}.
The experimental sensitivity is only limited by quantum noise in principle and it can probe tiny axion-photon coupling $g_{\text{a}\gamma} \lesssim 10^{-11} ~\text{GeV}^{-1}$ with axion mass range $m \lesssim 10^{-10} ~\text{eV}$ which is competitive with other experimental proposals.
Moreover, this new method can be highly advantageous compared with the conventional axion detectors, since it does not require superconducting magnets which often drive large cost.
Therefore, we expect that this method opens a new window to the axion dark matter research.

Inspired by these proposals using optical cavity, in this Letter we propose a new scheme to search for axion dark matter by using a linear Fabry-P\'erot cavity.
Linear optical cavities are used in the current and future gravitational wave detectors, such as Advanced LIGO (aLIGO)~\cite{Aasi2015}, 
Advanced Virgo~\cite{Acernese2014},
KAGRA~\cite{Somiya2012}, Einstein Telescope~\cite{Punturo2010}, Cosmic Explorer (CE)~\cite{Abbott2017}, and DECIGO~\cite{Kawamura2008}.
In this work, we explore the capabilities of these laser interferometers
to search for axion-like dark matter.
Remarkably, our new method enables the interferometers to probe axion-like dark matter during the gravitational wave observation run without loosing any sensitivity to gravitational waves.
It implies that we can exploit the cutting-edge laser facilities for axion-like dark matter search even without
constructing dedicated one from scratch.
We estimate the potential sensitivity to the axion-photon coupling with the parameter sets of gravitational wave observatories. 
Their sensitivities can overcome the current upper limit with broad axion mass range and put better bounds than the proposed axion experiments by several orders of magnitudes.
Note that although the gravitational wave detectors are discussed, we do not propose to measure gravitational waves. Our target is the phase velocity difference in circular-polarized photons and laser interferometer
is suitable for its detection. Thus, our proposal is complementary to the previous study of the gravitational waves sourced by axion~\cite{Brito:2017wnc}.

This Letter is organized as follows.
In the next section, we shortly derive the difference in the phase velocity of polarized photons in the presence of axion dark matter.
Then we present the scheme to detect it as polarization modulation of a linearly polarized light  using the Fabry-P\'erot cavity enhancing the modulation.
Next we describe the prospected sensitivity curves of axion-photon coupling with each gravitational wave interferometer.
Finally, we give a short discussion and conclude our result.
In this Letter, we set the natural unit $\hbar = c = 1$.

 
\section{Phase Velocity Modulation}

In this section, we briefly explain how the dispersion relations of two circular-polarized photons are modified in the 
presence of background axion field.
The axion-photon coupling is written as Chern-Simons interaction
\begin{equation}
\frac{g_{\text{a}\gamma}}{4} a(t) F_{\mu \nu} \tilde{F}^{\mu \nu} 
= g_{\text{a}\gamma} \dot{a}(t) \epsilon_{ijk} A_i \partial_j A_k + \text{(total derivative)}, \label{eq: 1}
\end{equation}
where the dot denotes the time derivative, $g_{\text{a}\gamma}$ is a coupling constant, $a(t)$ is the axion field value, and $A_\mu$ is 
the vector potential of the electromagnetic field strength $F_{\mu \nu} \equiv \partial_\mu A_\nu - \partial_\nu A_\mu$.
Its Hodge dual is defined as $\tilde{F}^{\mu\nu} \equiv \epsilon^{\mu \nu \rho \sigma}
F_{\rho\sigma}/2$, where $\epsilon^{\mu\nu\rho\sigma}$ is the Levi-Civita anti-symmetric tensor.
Regarding the gauge condition, we choose the temporal gauge $A_0 = 0$ and the 
Coulomb gauge $\partial_i A_i = 0$.
$A_i$ can be decomposed into two circular polarization modes in the Fourier space
\begin{equation}
A_i(t, \bm{x}) = \sum_{\lambda = \text{L}, \text{R}}\int\dfrac{\dd^3k}{(2\pi)^3}A_\lambda(t,\bm k)\,e^\lambda_i(\hat{\bm{k}})\,e^{i\bm{k}\cdot\bm{x}} \ ,
\end{equation}
where $\bm{k}$ is the wave number vector, the circular polarization vectors satisfy $e^\lambda_i(\hat{\bm{k}}) = e^{\lambda*}_i(-\hat{\bm{k}})$, $e^\lambda_i(\hat{\bm{k}})e^{\lambda'*}_i(\hat{\bm{k}}) = \delta^{\lambda\lambda'}$ and $i\epsilon_{ijm}k_je^{\text{L/R}}_m(\hat{\bm{k}}) = \pm k e^{\text{L/R}}_i(\hat{\bm{k}})$ 
($k \equiv |\bm{k}|$).
Here the index of L (R) corresponds to the upper (lower) sign of the double sign. Hereafter, we use the same notation in this Letter.
Then the equation of motion for $A_{\text{L/R}}(t,\bm k)$,
$\ddot{A}_{\text{L/R}}+\omega^2_{\text{L/R}}A_{\text{L/R}}=0$, acquires the modified dispersion relation due to the axion-photon coupling Eq.~\eqref{eq: 1},
$\omega_{\text{L/R}}^2=k^2\left(1\mp g_{\text{a}\gamma} \dot{a}/k\right)$.
This leads to the different phase velocities for the left and right polarization modes
\begin{align}
c_{\text{L/R}}^2 = 1 \mp \dfrac{g_{\text{a}\gamma}\dot{a}}{k} \ .
\label{CLR2}
\end{align}
Note that the momentum effect of axion dark matter here is irrelevant since it is non-relativistic.
Ignoring the cosmic expansion, the present axion dark matter is given by the periodic function
\begin{equation}
a(t) = a_0\cos(mt + \delta_\tau(t))
\label{axion oscillation}
\end{equation}
with the frequency of axion mass $f = m/(2\pi) \simeq 2.4 ~\text{Hz}~(m/10^{-14}~\text{eV})$.
The phase factor $\delta_\tau(t)$ can be regarded as a constant 
value within the coherent time scale
of axion dark matter, $\tau$, 
expressed as $\tau = 2\pi/(m v_\text{a}^2)$, where $v_\text{a}$ is 
axion dark matter velocity. Since the local velocity of dark matter is about $10^{-3}$, 
$\tau$ is estimated as
\begin{equation}
\tau \sim 1 \left( \frac{10^{-16} \text{ eV}}{m} \right) \text{ year}.
\end{equation}
Plugging Eq.~\eqref{axion oscillation} into \eqref{CLR2}, we obtain 
\begin{align}
c_\text{\text{L/R}}(t) &\simeq 1 \pm \delta c(t) 
\equiv 1 \pm \delta c_0 \sin(m t + \delta_\tau(t)), \label{eq:SoLofL} 
\end{align}
where $\delta c_0=g_{\text{a}\gamma}a_0m/(2k)$ is the maximum difference of the phase velocity, $c_0$ is the speed of light without background axion, and $\delta c_0\ll 1$ is used.
$\delta c_0$ is estimated as
\begin{align}
\delta c_0 \simeq 1.3 \times 10^{-24} \left( \frac{\lambda}{1550 \text{ nm}}\right) 
\left( \frac{g_{\text{a}\gamma}}{10^{-12} \text{ GeV}^{-1}}\right) .
\end{align}
Here we assumed the laser light with a wavelength $\lambda = 2\pi/k$ and used the present energy density value of axion dark matter around earth, $\rho_a = a_0^2m^2/2 \simeq 0.3 \text{ GeV/cm}^3$, which removes the dependence of $\delta c_0$ on $a_0 m$.

The key point is that according to the equation (\ref{eq:SoLofL}) 
one linearly polarized light (e.g. horizontal polarization, that is p-polarization) is polarization-modulated due to axion dark matter and the orthogonally polarized light
(e.g. vertical polarization, that is s-polarization) is produced as shown later.
Note that the linearly polarized light can be 
expressed by a superposition of two circularly polarized lights.
From the next section, we show that this polarization-modulation can be measured with linear cavities of gravitational wave experiments by using our proposed method.

\section{Axion search with a linear optical cavity}

\begin{figure}[htb]
  \centering
  \includegraphics[width=\columnwidth]{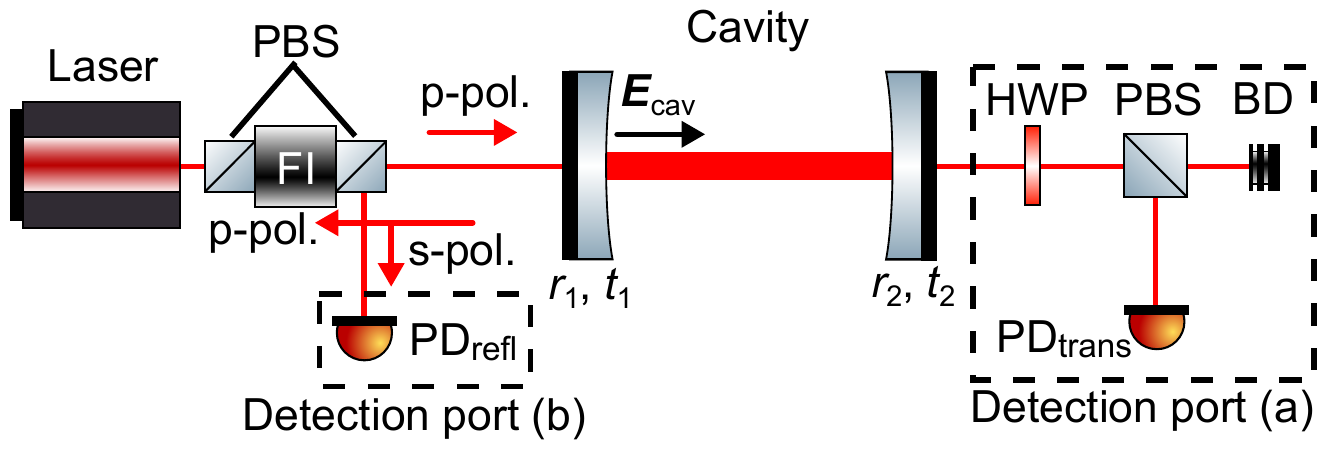}
  \caption{Schematic of experimental setup for axion search with a linear optical cavity. 
  FI, Faraday isolator; HWP, half wave plate; PBS, polarizing beam splitter; 
  PD, photodetector; BD, beam dump. Signal is detected in detection port (a) 
  and (b). Components for phase measurement are not shown. The polarization of incident 
  light is arbitrary if only it is linear polarization. Two PBSs in FI are placed 
  rotated by 45 degrees along the optical path.}
  \label{fig:Setup}
\end{figure}

In this section, we present how to detect the modulation of speed of light with linear optical cavities. The schematic setup of our proposed scheme is shown in figure \ref{fig:Setup}.
First, as a carrier wave, we input linearly-polarized monochromatic laser light with the angular frequency which corresponds to the wave number $k$.
Here, we consider p-polarized light as input light without loosing generality.
The cavity consists of the input and output mirrors whose amplitude reflectivities and transmissivities are represented by ($r_1$, $t_1$) and ($r_2$, $t_2$).
In this letter, we only consider axion mass range where $\tau$ is longer than the cavity storage time, $4\pi L\sqrt{r_1r_2}/(1-r_1 r_2)$~\cite{Yariv2007}.
In this condition, the axion can be treated as a coherent oscillator 
during the time when photon is interacting with the axion in the cavity.
When the cavity is kept to resonate with a phase measurement, such as Pound-Drever-Hall technique~\cite{Drever1983}, the beam is accumulated inside cavity and the signal, $\delta c$, is enhanced as explained later.
Then the signal is detected in detection port (a) or (b) as polarization modulation with polarizing optics. 
In detection port (a), the polarization of transmitted light from the cavity is slightly 
rotated by the half wave plate. Then, the photodetector ($\text{PD}_\text{trans}$) 
receives  s-polarized light generated by axion-photon coupling as a beatnote with faint (but much stronger than the signal) carrier wave, while most of the carrier light is transmitted by the polarizing beam splitter (PBS).
In detection port (b), the PD ($\text{PD}_\text{refl}$) receives signal reflected by the faraday isolator (FI) as a beatnote with faint carrier wave again.
In this case, the carrier wave is generated by non-ideal birefringence between the cavity and FI, such as input mirror substrate.
These two detection ports can be added without modifying the instrument for the phase measurement.

The signal, $\delta c$, is enhanced inside the cavity by the following mechanism. Here, we treat $\delta_\tau(t)$  as constant since we only consider the axion mass range where the axion oscillation coherent time is sufficiently longer than the storage time of the optical cavity.
The input p-polarized light is written as 
\begin{equation}
\bm{E}_\text{in}(t) = \bm{E}^\text{p}(t) = E_0 e^{i k t} 
\begin{pmatrix}
        \bm{e}^\text{L} & \bm{e}^\text{R}
\end{pmatrix}
\frac{1}{\sqrt{2}}
\begin{pmatrix}
    1 \\
    1
\end{pmatrix},\\
\end{equation}
where $\bm{E}^\text{p}(t)$ is the electric vector of p-polarized light, $\bm{e}^\text{L}$ and $\bm{e}^\text{R}$ are basis vectors of left-handed and 
right-handed laser light, respectively.
In the presence of background axion field, the electric vector propagation in the cavity in front of the front mirror is expressed as
\begin{gather}
\bm{E}_\text{cav}(t) =  t_1 E_0 e^{i k t} 
\begin{pmatrix}
        \bm{e}^\text{L} & \bm{e}^\text{R}
\end{pmatrix}
\sum^\infty_{n=1} A_n(t)
\frac{1}{\sqrt{2}}
\begin{pmatrix}
    1 \\
    1
\end{pmatrix}, \\
\begin{cases}
    \begin{split}
    A_{n+1}(t) \equiv &  A_{n}(t) R_1 T(t-2L(n-1)) \\
    &  \ \ \ \times R_2 T(t-2L(n-1/2))  
    \end{split}& (n \ge 1) \\
    A_1 = 1    
\end{cases},
\end{gather}
where $L$ is cavity length, $T(t)$ is transfer matrix for one-way translation,
\begin{align}
T(t) &\equiv 
\begin{pmatrix}
        e^{-i\phi^\text{L}(t)} & 0 \\
        0 & e^{-i\phi^\text{R}(t)}
\end{pmatrix}, \\
\phi^\text{\text{L/R}}(t) &\equiv k L 
\mp k \int^t_{t-L} \delta c(t^\prime) dt^\prime, 
\end{align}
and $R_i$ is reflection matrix for circularly-polarized lights,
\begin{equation}
R_i \equiv
\begin{pmatrix}
        0 & -r_i \\
        -r_i & 0
\end{pmatrix}
 \ \ \ (i = 1, 2). \label{eq:ReflMat}
\end{equation}
Sign flipping in eq. (\ref{eq:ReflMat}) is the main difference from the modeling in \cite{DeRocco2018, Obata2018, Liu2018}. Here $A_n(t) \ \ (n\ge 2)$ is given by,
\begin{equation}
A_n(t) = (r_1 r_2)^{n-1}
\begin{pmatrix} \displaystyle
        A^{11}_n(t) & 0 \\
        0 & A^{22}_n(t)
\end{pmatrix}
\end{equation}
with
\begin{align}
&A^{11/22}_n \equiv \exp\bigg[ -i k \bigg\{ 2 L(n-1) 
\notag\\
&\pm \sum^{n-1}_{j=1}\bigg( \int^{t-2L(j-1)}_{t-2L(j-1/2)} 
- \int^{t-2L(j-1/2)}_{t-2Lj} \bigg) 
\delta c(t^\prime) dt^\prime\bigg\}\bigg],
\label{eq:An_of_t}
\end{align}
where the 11 and 22 component of $A_n$ corresponds to the upper and lower sign of the flipped sign, respectively.
When resonance condition of the linear cavity, $2k L = 2 \pi l \ \ (l\in\mathbb{N})$, 
is met, $A^{11}_n$ and $A^{22}_n$ is also denoted as 
\begin{align}
A^{11/22}_n = \exp\bigg[ \mp i 
        &k \int^\infty_{-\infty} \tilde{\delta c}(m)\frac{1}{m}
        \tan\left(\frac{m L}{2} \right) 
        \notag\\
        &\ \ \times\left(1-e^{i 2m L(n-1)}\right)
        e^{im t} \frac{dm}{2\pi}\bigg],
\end{align}
where we transformed $\delta c(t)$ in Fourier space, 
$\delta c(t) = \int^\infty_{-\infty} \tilde{\delta c}(m)e^{i m t} \frac{dm}{2\pi}$.
Consequently, the electronic field in the cavity is written as, 
\begin{align}
\bm{E}_\text{cav}(t) =& \frac{t_1 E_0e^{i k t}}{1-r_1r_2}
\begin{pmatrix}
        \bm{e}^\text{L} & \bm{e}^\text{R}
\end{pmatrix} \nonumber \\
&  \ \ \ \ \times \begin{pmatrix}
      1 + i \delta \phi(t) & 0 \\
      0 & 1 - i \delta \phi(t)
\end{pmatrix}
\frac{1}{\sqrt{2}}
\begin{pmatrix}
      1 \\
      1
\end{pmatrix} \\
=& \frac{t_1}{1-r_1r_2}\left[\bm{E}^\text{p}(t) 
- \delta \phi(t)\bm{E}^\text{s}(t)\right],
\end{align}
where $\bm{E}^\text{s}$ are electric vectors of s-polarized light,
\begin{equation}
 \delta \phi(t) \equiv \int^\infty_{-\infty} 
 \tilde{\delta c}(m) H_\text{a}(m) e^{im t} \frac{dm}{2\pi},
\end{equation}
and $H_\text{a}(m)$ is a response function of cavity,
\begin{equation}
 H_\text{a}(m) \equiv i \frac{k}{m} 
 \frac{4 r_1r_2\sin^2\left(\frac{m L}{2}\right)}
 {1 - r_1 r_2 e^{-i2m L}}  \left(-e^{-imL}\right) \label{eq:Ha}.
\end{equation}
Equation (\ref{eq:Ha}) indicates that the signal is enhanced in proportion to 
$r_1 r_2/(1-r_1 r_2)$ at 
$m = \pi/L$, which corresponds to the free spectral range, i.e. the frequency separation of the longitudinal mode of the cavity~\cite{Yariv2007}. The peak sensitivity can be enhanced by increasing the mirror reflectivity although finesse, $\pi\sqrt{r_1 r_2}/(1-r_1 r_2)$, is limited to be lower than $10^6$ due to the dispersion of the dark matter~\cite{Millar2017}. In addition, $H_\text{a}(m) \propto 1/m$ at $mL= \pi(2N-1) \ (N\in\mathbb{N})$ since the axion effect on the photons in the cavity is cancelled out except for the last half of the axion oscillation when the axion oscillation period is shorter than the photon storage time of the cavity ~\cite{Yariv2007}. 
In low mass range ($mL \ll 1$), $H_\text{a}(m) \propto m$ since the axion effect is cancelled on going and returning way due to eq. (\ref{eq:ReflMat}).



\section{Sensitivity to the axion-photon coupling}

In this section, we estimate the potential sensitivity of the linear cavity to axion-photon coupling. Here, only shot noise which is caused by vacuum fluctuation of electric field, 
$E_\text{vac}(t)$, is considered in a similar way to shot-noise estimation of gravitational wave detectors~\cite{Kimble2001a}. In each detection port, the electric field received by 
photodetector is expressed as 

\begin{align}
\bm{E}_\text{PD}(t) =& \left[\sqrt{\mathcal{T}_j}
\left(\alpha - \delta \phi(t)\right)
 + \frac{E_\text{vac}(t)}{E_0} \right]\bm{E}^\text{s}(t) \ \ (j=1, 2), \\
\sqrt{\mathcal{T}_j} \equiv& \frac{t_1t_j}{1-r_1r_2},
\end{align}
where $\alpha$ $(|\alpha| \ll 1)$ is the polarization mixing angle 
introduced by the half wave plate (instrumental birefringence) and $j = 2 \ (1)$ for 
the detection port (a) ((b)). Here, we neglect the second and higher order of $|\alpha|$. Note that $|\alpha|$ is much larger than $|\delta \phi(t)|$ and $|E_\text{vac}(t)|$.
The detected power is
\begin{align}
&P_\text{PD}(t)   \propto |\bm{E}_\text{PD}(t)|^2  \nonumber \\
&\simeq  \alpha \sqrt{\mathcal{T}_j}  E_0^2\left[\alpha\sqrt{\mathcal{T}_j}
-2\sqrt{\mathcal{T}_j}\delta \phi(t) + 2\frac{E_\text{vac}(t)}{E_0}\right],
\label{eq:SignalPower}
\end{align}
where 
the second order and cross terms of $\delta \phi(t)$ and $E_\text{vac}(t)$ are ignored.
We can estimate the sensitivity by comparing the second and third terms of equation (\ref{eq:SignalPower}) which are time-dependent.
The second and third term corresponds to signal and shot noise, respectively. The one-sided linear spectrum of shot noise equivalent to $\tilde{\delta c}(m)$, 
$\sqrt{S_\text{shot}(m)}$, is obtained by considering the ratio of the noise term to the signal term,
\begin{equation}
\sqrt{S_\text{shot}(m)} = \frac{\sqrt{\frac{k}{2 P_0}}}
{\sqrt{\mathcal{T}_j}|H_\text{a}(m)|},
\label{eq:Shotnoise}
\end{equation}
where $P_0$ is incident power.
Here, we used $E_0 = \sqrt{2 P_0/k}$ and the one-sided spectrum of vacuum fluctuation is 
unity~\cite{Kimble2001a}.
In this Letter, the electric field has dimensions of [$\sqrt{\text{Hz}}$] as in  \cite{Kimble2001a}.
According to the equation (\ref{eq:Shotnoise}), if the cavity is over-coupled, i.e. $t_1 > t_2$, 
detection port (b) is better.
On the other hand, detection port (a) is effective for the critical-coupled cavity, i.e. $t_1 = t_2$, since there is no carrier wave in the reflection port under the critical coupling condition.

If the sensitivity is limited by shot noise, the signal-to-noise ratio (SNR) for $\delta c_0$ is improved with measurement time, $T_\text{obs}$. The improvement 
depends on whether $T_\text{obs}$ is larger than the coherent time of axion oscillation, $\tau$, or not \cite{Budker2014}:
\begin{equation}
\text{SNR} = 
  \begin{cases}
    \frac{\sqrt{T_\text{obs}}}{2\sqrt{S_\text{shot}(m)}}\delta c_0
    & (T_\text{obs} \lesssim \tau) \\
    \frac{(T_\text{obs} \tau)^{1/4}}{2\sqrt{S_\text{shot}(m)}}\delta c_0
    & (T_\text{obs} \gtrsim \tau)
  \end{cases} .
\end{equation}
We can find the detectable value of $\tilde{\delta c}(m)$ which
sets the SNR to unity
\begin{equation}
\delta c_0 \simeq
  \begin{cases}
    \frac{2}{\sqrt{T_\text{obs}}}\sqrt{S_\text{shot}(m)} & (T_\text{obs} \lesssim \tau) \\
    \frac{2}{(T_\text{obs} \tau)^{1/4}}\sqrt{S_\text{shot}(m)} & (T_\text{obs} \gtrsim \tau)
  \end{cases} .
\end{equation}
Finally, this is translated into the sensitivity to $g_{\text{a}\gamma}$ as
\begin{align}
g_{\text{a}\gamma}(m) &\simeq 1.5\times10^{12}\,{\rm GeV}^{-1} 
\left( \frac{1550 \text{ nm}}{\lambda} \right) 
\notag\\&\quad\times
  \begin{cases}
    \sqrt{\frac{S_\text{shot}(m)}{T_\text{obs}}}  
    & (T_\text{obs} \lesssim \tau) \\
    \sqrt{\frac{S_\text{shot}(m)}{\sqrt{T_\text{obs} \tau}}} 
    & (T_\text{obs} \gtrsim \tau)
  \end{cases} .
\end{align}

\begin{figure}[htb]
  \centering
  \includegraphics[width=\columnwidth]{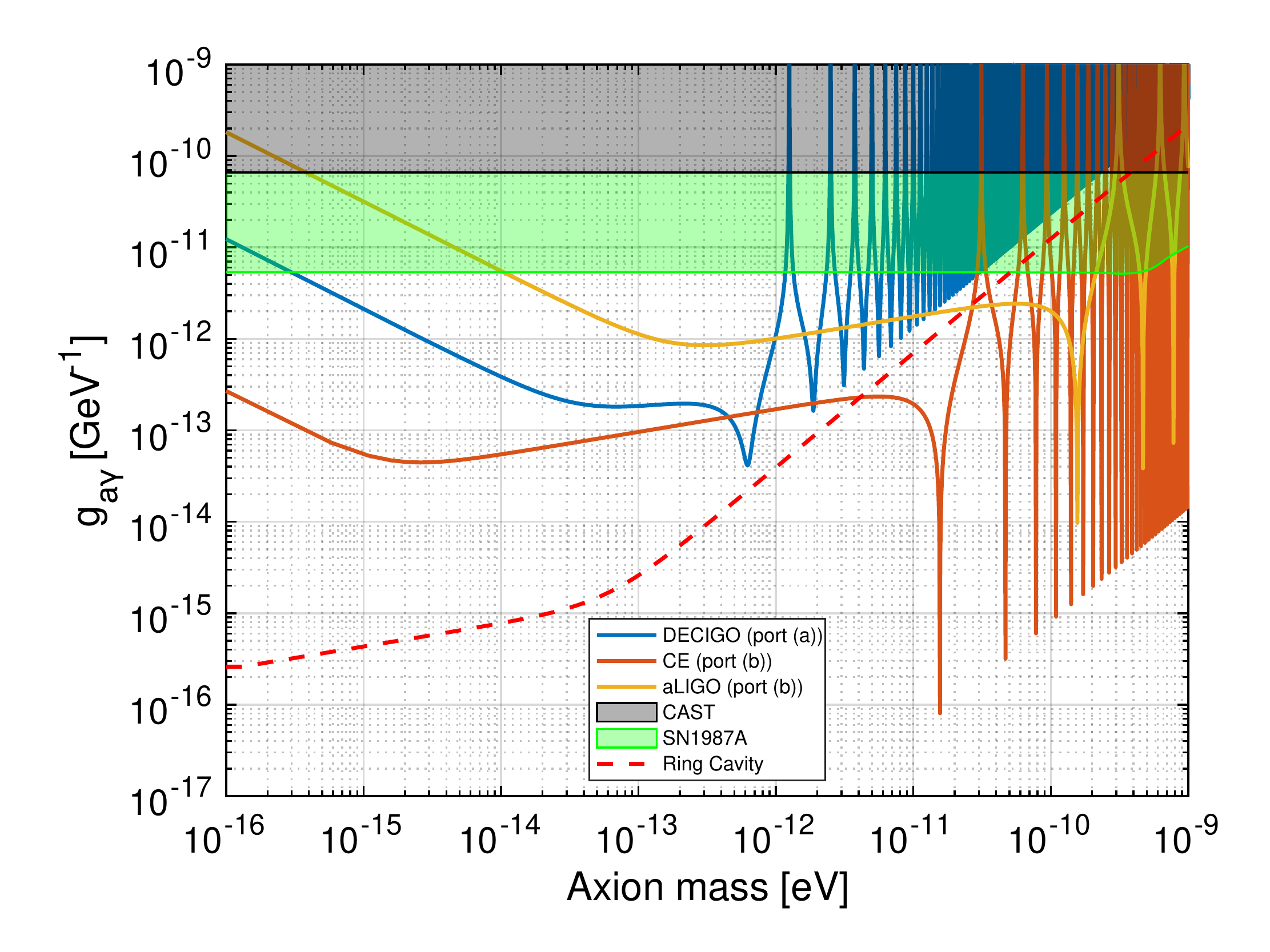}
  \caption{Sensitivity comparison of the several parameter sets shown in table 
  \ref{tab:ITFparameters}. Although the higher mass range seems to be filled, 
  they have sensitivity peaks at mass of $m=\pi(2N-1)/L \ (N\in\mathbb{N})$.
  The gray and green band express the current limit provided by 
  CAST~\cite{CASTCollaboration2005} and the cosmic ray observations of SN1987A~\cite{Payez2015}.
  The red dashed line is a 
  sensitivity curve of one proposed experiment using optical ring cavity with optimistic 
  parameters~\cite{Obata2018}.}
  \label{fig:SensComp}
\end{figure}

Figure \ref{fig:SensComp} shows the shot-noise limited sensitivities 
to $g_{\text{a}\gamma}$ with our scheme.
Here, we adopted the experimental parameter sets used or planned by gravitational wave detectors (specifically, DECIGO~\cite{Kawamura2008}, 
CE~\cite{Abbott2017}, and aLIGO~\cite{Aasi2015}) as shown in table \ref{tab:ITFparameters}.
We also assume $T_\text{obs} = 1$ year and $r_i^2 + t_i^2 = 1$.
Note that detection port (a) is used for DECIGO-like 
detector and port (b) is used for CE- and aLIGO-like detectors.
All gravitational wave detectors have a sensitive mass range similar to some proposed experiment such as IAXO~\cite{Vogel2013} and ABRACADABRA~\cite{Kahn2016}.
In all parameter sets, the upper limit provided by CAST ~\cite{CASTCollaboration2005} can be improved.
Especially, the CE-like detector can overcome the CAST limit by three 
orders of magnitude in broad mass range around between 
$4\times10^{-16}$ and $1\times10^{-13}$ eV.
At the most sensitive mass $m=1.563 \times 10^{-11}$ eV, the improvement from CAST limits is about 6 orders of magnitude although QCD axions cannot be detected. 

\begin{table*}[htb]
\centering
 \caption{Parameters of considered gravitational wave detectors. Note that $P_0$ is the input beam power to front mirror enhanced by the power recycling cavity for aLIGO-like and CE-like detector~\cite{Drever1983a}.}
 \label{tab:ITFparameters}
 \begin{tabular}{c|c|c|c|c}
  Similar detector      &       $L$     [m] &   $P_0$ [W]       &       $\lambda$ [$\times 10^{-9}$ m]    &       $(t_1^2, t_2^2)$ [ppm]  
  \\ \hline
  DECIGO~\cite{Kawamura2008}    &       $10^6$  &       5       &       515     &       ($3.1 \times 10^5$, $3.1 \times 10^5$)        \\ \hline
  CE~\cite{Abbott2017} &       $4\times10^4$   &       600     &       1550    &       ($1.2\times10^3$, 5)      \\ \hline
  aLIGO~\cite{Aasi2015}        &       $4\times10^3$   &       2600    &       1064    &       ($1.4\times10^4$, 5)
 \end{tabular}
\end{table*}

It is worth noting that in our scheme the displacement noise such as the vibration of mirrors or the gravitational wave signal itself does not become manifest unlike a gravitational wave detector.
This is because the displacement noises and gravitational waves make the same phase shift in the two circularly-polarized lights propagating in the same path 
and this phase shift is cancelled in the measurement of the phase difference between two polarized lights.
A major technical noise source in our scheme is a roll motion of the mirrors 
which would generate relative phase shift in the two polarized lights through birefringence of the mirror coating. The effect of the substrate birefringence is relatively small since the signal is enhanced in the cavity. When the laser polarization and coating axis are almost aligned, the noise spectrum is expressed as $\sqrt{S_\text{roll}} \simeq \delta \alpha~\lambda~\theta_\text{bi}/L$, where $\delta \alpha$ is roll motion spectrum and $\theta_\text{bi}$ is a coating birefringence. In aLIGO case, $\theta_\text{bi} \simeq 10 \text{ $\mu$rad}$~\cite{Winkler1994}. The seismic motion make $\delta \alpha < 10^{-11}$ rad/Hz$^{1/2}$ for $m>10^{-14}$ eV if we conservatively assume that coupling from vertical to roll motion is unity ~\cite{Aston2012a}. Thus, $\sqrt{S_\text{roll}} < 3\times10^{-26}$ 1/Hz$^{1/2}$, which is smaller than shot noise level. In DECIGO and CE, the roll motion of the mirror would be small since they would be in space or underground site while aLIGO is on the ground.

In order to apply our method to the real gravitational detector, some optics are
added for detection port and there exist constructional problems.
The approach to detect the signal in detection port (b) is not quite simple because there have been equipped several apparatuses, such as a beam splitter, a signal recycling mirror~\cite{Meers1988, Mizuno1993}, and so on, between the front mirror and FI.
In principle, the axion signal can be extracted behind the signal recycling mirror as with 
the gravitational wave signal readout~\cite{Fricke2012}.
More practical issues will be investigated in future work. 

\section{Conclusion}

We developed the experimental scheme to search for axion-like dark matter with the optical linear cavity used in gravitational wave detectors.
Our experiment measures the production of the linear polarization component opposite to the intrinsic polarization of the incident laser beam caused by the axion-photon coupling.
The experimental sensitivity is in principle limited only by quantum shot noise, and other kind of technical disturbances 
are irrelevant.
We estimated the potential sensitivity of detectors to the axion-photon coupling in a broad mass range $10^{-16}\text{~eV} \lesssim m \lesssim 10^{-9}\text{~eV}$ with the experimental parameters of existing gravitational wave detector projects, such as DECIGO, CE and aLIGO.
As a result, we found that their sensitivities can reach beyond the current limit of CAST~\cite{CASTCollaboration2005} with a wide axion mass range and can be competitive with other experimental proposals which were recently suggested ~\cite{DeRocco2018, Obata2018, Liu2018}. 
Remarkably, our new scheme for axion-like dark matter search can be performed with a minor modification of the gravitational wave detector and coexist with its observation run for gravitational waves.
We expect that this scheme becomes a new approach to search for axion dark matter.

\section{Acknowledgement}
\label{Acknowledgement}

In this work, KN, YM and TF are supported by the JSPS KAKENHI Grant No.~JP17J01176, JSPS Grant-in-Aid for Scientific Research (B) No.~18H01224 and Grant-in-Aid for JSPS Research Fellow No.~17J09103, respectively.




\end{document}